\documentclass[letterpaper, 10 pt, conference]{ieeeconf}  

\IEEEoverridecommandlockouts                              
\usepackage{amsmath} 
\usepackage{epsfig}
\usepackage{graphicx}
\usepackage[hang]{caption2}
\usepackage[normalsize,it,hang]{subfigure}
\usepackage[latin1]{inputenc}
\usepackage[greek,english]{babel}
\usepackage{longtable}
\usepackage{booktabs}

\newtheorem{remark}{Remark}[section]

\title{\LARGE \bf
Delta Hedging in Financial Engineering: \\ Towards a Model-Free
Approach}

\author{Michel {\sc{Fliess}}, C\'edric {\sc{Join}}
\thanks{Michel {\sc{Fliess}} is with INRIA-ALIEN \& LIX
(CNRS, UMR~7161), \'Ecole polytechnique, 91128 Palaiseau, France.
\newline {\tt \small Michel.Fliess@polytechnique.edu}}
\thanks{C\'{e}dric {\sc{Join}} is with INRIA-ALIEN \& CRAN (CNRS, UMR 7039), Nancy-Universit\'e,
BP 239, 54506 Vand\oe uvre-l\`es-Nancy, France. \newline {\tt \small
cedric.join@cran.uhp-nancy.fr}} }

\begin{document}

\maketitle
\thispagestyle{empty}
\pagestyle{empty}

\begin{abstract}
Delta hedging, which plays a crucial r\^{o}le in modern financial
engineering, is a tracking control design for a ``risk-free''
management. We utilize the existence of trends in financial time
series (Fliess M., Join C.: \emph{A mathematical proof of the
existence of trends in financial time series}, Proc. Int. Conf.
Systems Theory: Modelling, Analysis and Control, Fes, 2009. Online:
{\tt http://hal.inria.fr/inria-00352834/en/}) in order to propose a
model-free setting for delta hedging. It avoids most of the
shortcomings encountered with the now classic Black-Scholes-Merton
framework. Several convincing computer simulations are presented.
Some of them are dealing with abrupt changes, {\it i.e.}, jumps.

\end{abstract}

\section{Introduction}
\emph{Delta hedging}, which plays an important r\^{o}le in financial
engineering (see, {\it e.g.}, \cite{taleb} and the references
therein), is a tracking control design for a ``risk-free''
management. It is the key ingredient of the famous
Black-Scholes-Merton (BSM) partial differential equation
(\cite{bs,merton}), which yields option pricing formulas. Although
the BSM equation is nowadays utilized and taught all over the world
(see, {\it e.g.}, \cite{hull,wilmott}), the severe assumptions,
which are at its bottom, brought about a number of devastating
criticisms (see, {\it e.g.},
\cite{derman,haug,ht,herlin,mandelbrot,swan,walter} and the
references therein), which attack the very basis of modern financial
mathematics, and therefore of delta hedging.

We introduce here a new dynamic hedging, which is influenced by
recent advances in {\em model-free} control
(\cite{esta,malo}),\footnote{See, {\it e.g.}, \cite{edf} for a most
convincing application.} and bypass the shortcomings due to the BSM
viewpoint:
\begin{itemize}
\item In order to avoid the study of the precise probabilistic nature of the
fluctuations (see the comments in \cite{trend,malo-sm}, and in
\cite{gol}), we replace the various time series of prices by their
{\em trends} \cite{trend}, like we already did for redefining the
classic beta coefficient \cite{cogis}.
\item The control variable satisfies an elementary algebraic
equation of degree $1$, which results at once from the {\em dynamic
replication} and which, contrarily to the BSM equation, does not
need cumbersome final conditions.
\item No complex calibrations of various coefficients are required.
\end{itemize}

\begin{remark}
Connections between mathematical finance and various aspects of
control theory has already been exploited by several authors (see,
{\it e.g.}, \cite{bernhard,pham,primbs} and the references therein).
Those approaches are however quite far from what we are doing.
\end{remark}
Our paper\footnote{See \cite{manuscript} for a first draft.} is
organized as follows.  The theoretical background is explained in
Section \ref{setting}. Section \ref{simulations} displays several
convincing numerical simulations which
\begin{itemize}
\item describe the behavior of $\Delta$ in ``normal'' situations,
\item suggest new control strategies when abrupt changes, {\it i.e.},
jumps, occur, and are forecasted via techniques from \cite{abrupt}
and \cite{malo-sm,cogis}.
\end{itemize}
Some future developments are listed in Section \ref{conclusion}.

\section{The fundamental equations}\label{setting}

\subsection{Trends and quick fluctuations in financial time
series}See \cite{trend}, and \cite{malo-sm,cogis}, for the
definition and the existence of {\em trends} and {\em quick
fluctuations}, which follow from the Cartier-Perrin theorem
\cite{cartier}.\footnote{The connections between the
Cartier-Perrin-theorem (see \cite{lobry} for an introductory
explanation) and {\em technical analysis} (see, {\it e.g.},
\cite{bechu,kabbaj,kirk}) are obvious (see \cite{trend} for
details).} Calculations of the trends and of its derivatives are
deduced from the denoising\footnote{The Cartier-Perrin theorem
permits to give a new definition of {\em noises} in engineering
\cite{bruit}.} results in \cite{nl,mboup} (see also \cite{garcia}),
which extend the familiar {\em moving average} techniques in
technical analysis (see, {\it e.g.}, \cite{bechu,kabbaj,kirk}).

\subsection{Dynamic hedging}
\subsubsection{The first equation}
Let $\Pi$ be the value of an elementary portfolio of one long
option position $V$ and one short position in quantity $\Delta$ of
some underlying $S$:
\begin{equation}\label{Pi}
\Pi = V - \Delta S
\end{equation}
Note that $\Delta$ is the control variable: the underlying asset is
sold or bought. The portfolio is {\em riskless} if its value obeys
the equation
$$
d \Pi = r(t) \Pi dt
$$
where $r(t)$ is the risk-free rate interest of the equivalent amount
of cash. It yields
\begin{equation}\label{val}
\Pi (t) = \Pi (0) \exp \int_{0}^{t} r({\tau}) d \tau
\end{equation}
Replace Equation \eqref{Pi} by
\begin{equation}\label{Pit}
\Pi_{\tiny{\rm trend}} = V_{\tiny{\rm trend}} - \Delta S_{\tiny{\rm
trend}}
\end{equation}
and Equation \eqref{val} by
\begin{equation}\label{valt}
\Pi_{\tiny{\rm trend}} = \Pi_{\tiny{\rm trend}} (0) \exp
\int_{0}^{t} r({\tau}) d \tau
\end{equation}
Combining Equations \eqref{Pit} and \eqref{valt} leads to the
tracking control strategy
\begin{equation}\label{hedging}
\boxed{\Delta = \frac{V_{\tiny{\rm trend}} - \Pi_{\tiny{\rm trend}} (0)
e^{\int_{0}^{t} r({\tau}) d \tau}}{S_{\tiny{\rm trend}}}}
\end{equation}
We might again call {\em delta hedging} this strategy, although it
is of course an approximate dynamic hedging via the utilization of
trends.

\subsubsection{Initialization} In order to implement correctly
Equation \eqref{hedging}, the initial values $\Delta (0)$ and
$\Pi_{\tiny{\rm trend}} (0)$ of $\Delta$ and $\Pi_{\tiny{\rm
trend}}$ have to be known. This is achieved by equating the
logarithmic derivatives at $t = 0$ of the right handsides of
Equations \eqref{Pit} and \eqref{valt}. It yields
\begin{equation}\label{0}
\boxed{\Delta (0) = \frac{\dot{V}_{\tiny{\rm trend}} (0) - r(0)
V_{\tiny{\rm trend}} (0)}{\dot{S}_{\tiny{\rm trend}} (0) - r(0)
S_{\tiny{\rm trend}} (0)}}
\end{equation}
and
\begin{equation}\label{00}
\boxed{\Pi_{\tiny{\rm trend}} (0) = V_{\tiny{\rm trend}} (0) - \Delta (0)
S_{\tiny{\rm trend}} (0)}
\end{equation}

\begin{remark}
Let us emphasize once more that the derivation of Equations
\eqref{hedging}, \eqref{0} and \eqref{00} does not necessitate any
precise mathematical description of the stochastic process $S$ and
of the volatility. The numerical analysis of those equations is
moreover straightforward.
\end{remark}

\begin{remark}
The literature seems to contain only few other attempts to define
dynamic hedging without having recourse to the BSM machinery (see,
{\it e.g.}, \cite{hf,gen,hut} and the references therein).
\end{remark}

\begin{remark}
Our dynamic hedging bears some similarity with \emph{beta hedging}
\cite{jorion}, which will be analyzed elsewhere.
\end{remark}

\subsection{A variant}\label{variant}
When taking into account variants like the {\em cost of carry} for
commodities options (see, {\it e.g.}, \cite{wilmott}), replace
Equation \eqref{Pit} by
\begin{equation*}\label{carry}
d\Pi_{\tiny{\rm trend}} = dV_{\tiny{\rm trend}} - \Delta
dS_{\tiny{\rm trend}} + q \Delta S_{\tiny{\rm trend}} dt
\end{equation*}
where $qSdt$ is the amount required during a short time interval
$dt$ to finance the holding. Combining the above equation with
$$d\Pi_{\tiny{\rm trend}} = r \Pi_{\tiny{\rm trend}}(0) \left( \exp
\int_{0}^{t} r(\tau) d \tau \right) dt$$ yields
\begin{equation*}\label{carry}
\Delta = \frac{{\dot V}_{\tiny{\rm trend}} - r \Pi_{\tiny{\rm
trend}}(0) \left( \exp \int_{0}^{t} r(\tau) d \tau \right)  }{{\dot
S}_{\tiny{\rm trend}} - q S_{\tiny{\rm trend}}}
\end{equation*}
The derivation of the initial conditions $\Delta (0)$ and
$\Pi_{\tiny{\rm trend}}(0)$ remains unaltered.

\section{Numerical simulations}\label{simulations}


\subsection{Two examples of delta hedging}
\begin{figure*}[htb]
\begin{center}\subfigure[\scriptsize Underlying asset: daily values
of the CAC from 28 April 2000 until 18 September
2009]{\includegraphics[scale=.5]{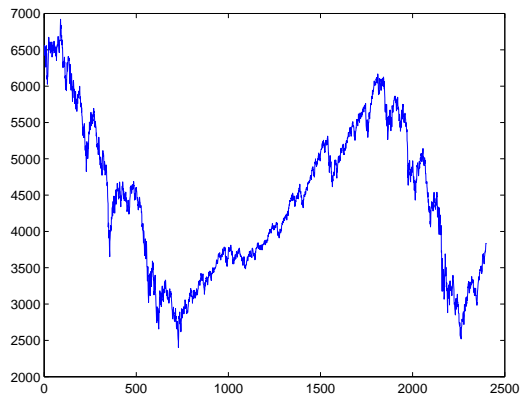}}\end{center}
\subfigure[\scriptsize Option: CFU9PY3500 daily prices from 9 May
2009 until 18 September 2009]{\includegraphics[scale=.5]{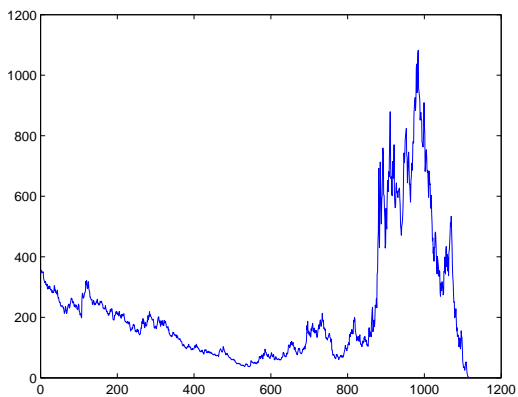}}
\subfigure[\scriptsize Option: CFU9CY3500 daily prices from 9 May
2009 until 18 September 2009 ]{\includegraphics[scale=.5]{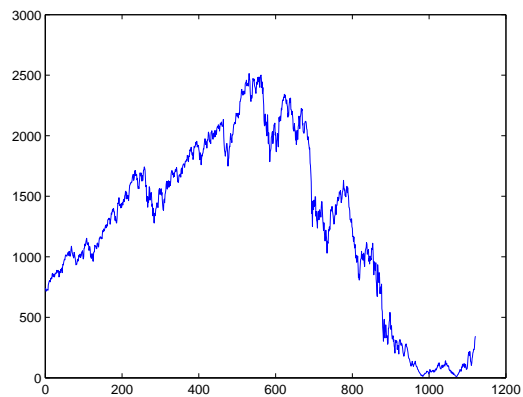}}
\caption{Daily data}\label{tout}
\end{figure*}

\begin{figure*}[htb]
\center\subfigure[\scriptsize Underlying asset: daily values during the
last 223 days, and trend (- -)]{\includegraphics[scale=.5]{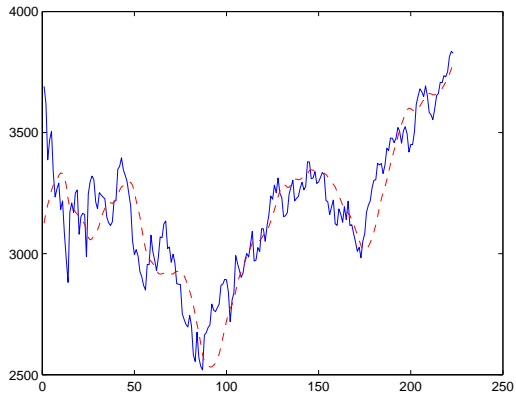}}
\subfigure[\scriptsize Option: daily values during the last 223
days, and trend  (- -)]{\includegraphics[scale=.5]{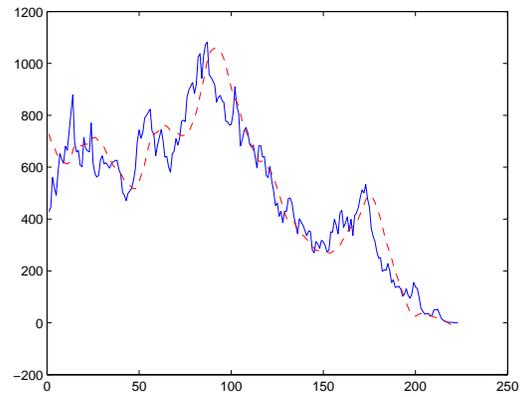}}
\center\subfigure[\scriptsize  Daily interest rate
$r$]{\includegraphics[scale=.5]{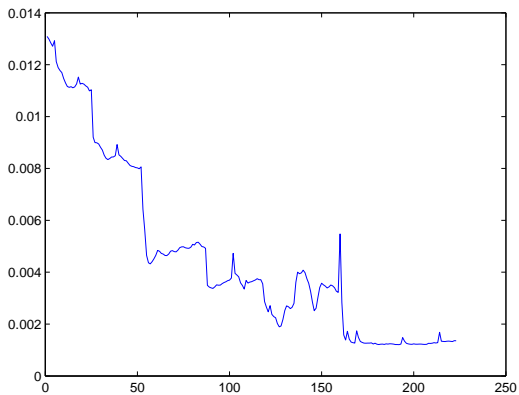}}
\subfigure[\scriptsize $\Delta$ tracking]{\includegraphics[scale=.5]{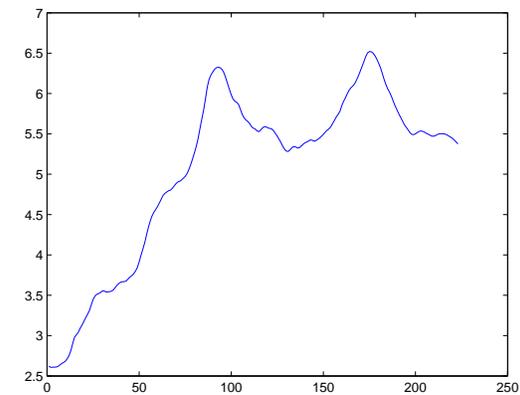}}
\caption{Example 1: CFU9PY3500 }\label{zoom1}
\end{figure*}

\begin{figure*}[htb]
\center\subfigure[\scriptsize Underlying asset: values during the last 223
days, and trend (- -)]{\includegraphics[scale=.5]{S.eps}}
\subfigure[\scriptsize Option: values during the last 223 days, and
trend (- -)  (- -)]{\includegraphics[scale=.5]{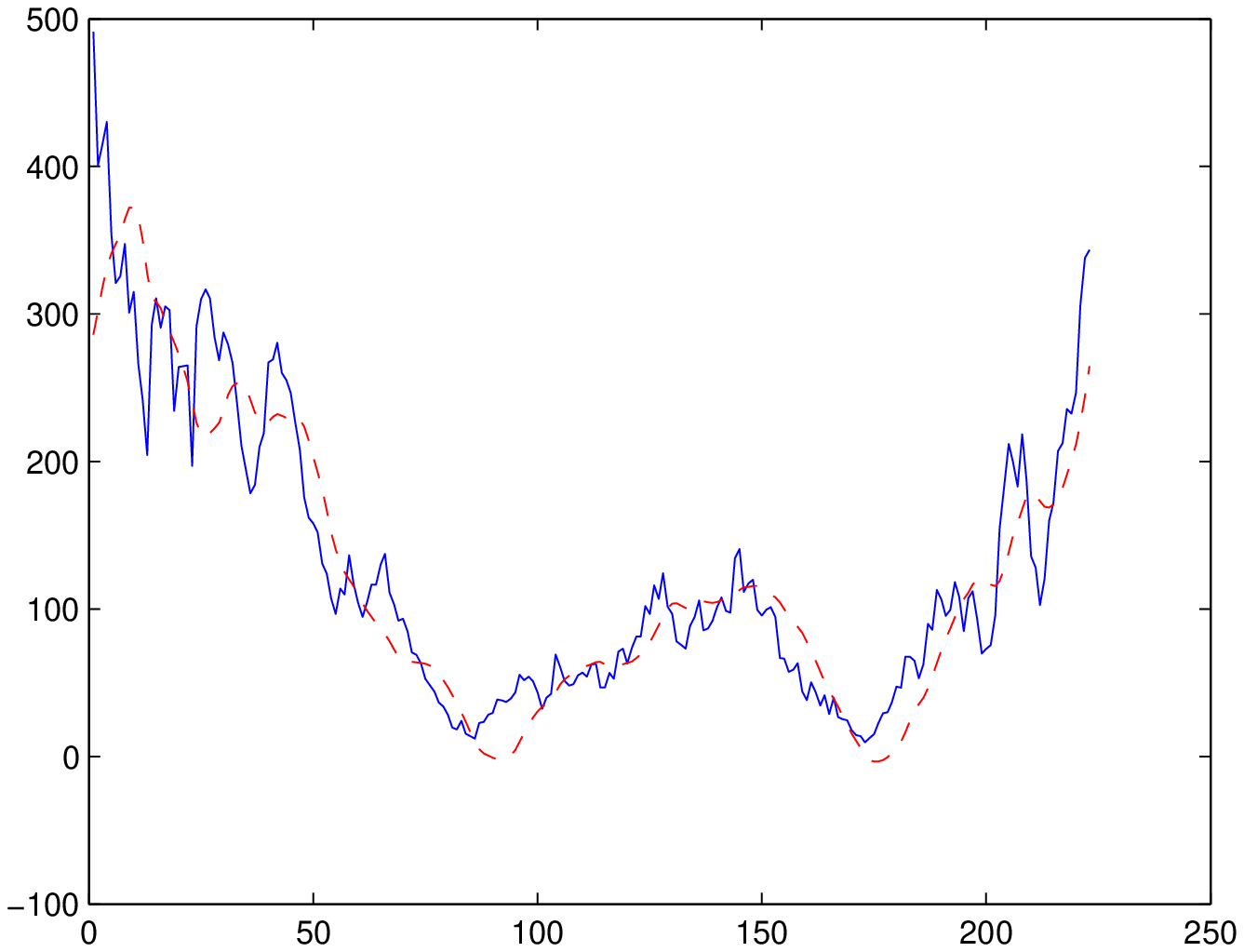}}
\center\subfigure[\scriptsize  Daily interest rate
$r$]{\includegraphics[scale=.5]{tau.eps}}
\subfigure[\scriptsize $\Delta$ tracking]{\includegraphics[scale=.5]{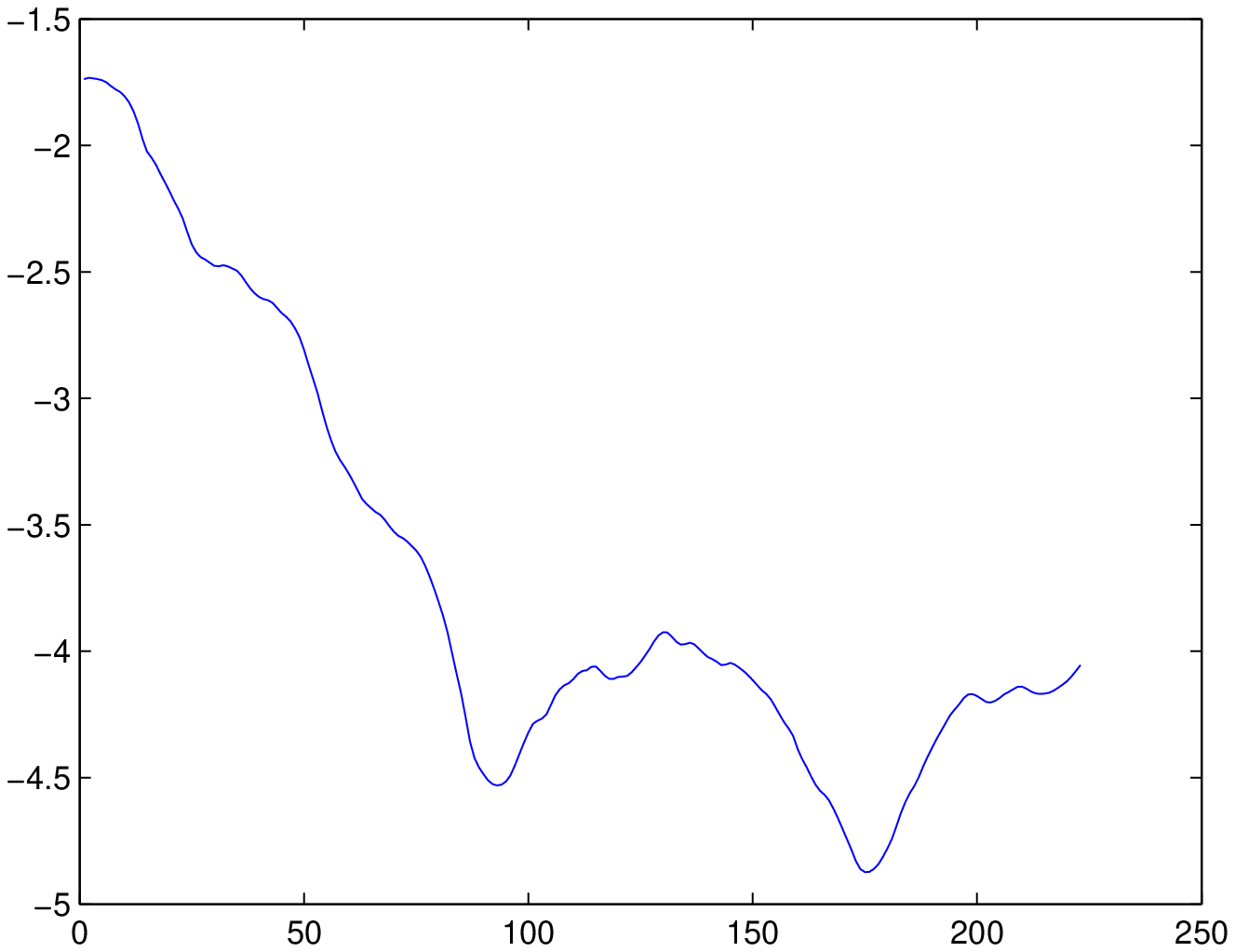}}
\caption{Example 2: CFU9CY3500 }\label{zoom2}
\end{figure*}

Take two derivative prices: one put (CFU9PY3500) and one call
(CFU9CY3500). The underlying asset is the CAC 40. Figures
\ref{tout}-(a), \ref{tout}-(b) and \ref{tout}-(c) display the daily
closing data. We focus on the 223 days before September 18$^{th}$,
2009. Figures \ref{zoom1}-(a) and \ref{zoom1}-(b) (resp.
\ref{zoom2}-(a) and \ref{zoom2}-(b)) present the stock prices and
the derivative prices during this period, as well as their
corresponding trends. Figure \ref{zoom2}-(c) shows the daily
evolution of the risk-free interest rate, which yields the tracking
objective. The control variable $\Delta$ is plotted in Figure
\ref{zoom2}-(d).
%


\subsection{Abrupt changes}
\subsubsection{Forecasts}
We assume that an abrupt change, {\it i.e.}, a jump, is preceded by
``unusual'' fluctuations around the trend, and further develop
techniques from \cite{abrupt}, and from \cite{malo-sm,cogis}. In
Figure \ref{anti}-(a), which displays forecasts of abrupt changes,
the symbols $o$ indicate if the jump is upward or downward.


%

\subsubsection{Dynamic hedging}
\begin{figure*}[htb]
\center\subfigure[\scriptsize Underlying (--), trend (- -),
prediction of abrupt change locations (l) and their directions
(o)]{\includegraphics[scale=.5]{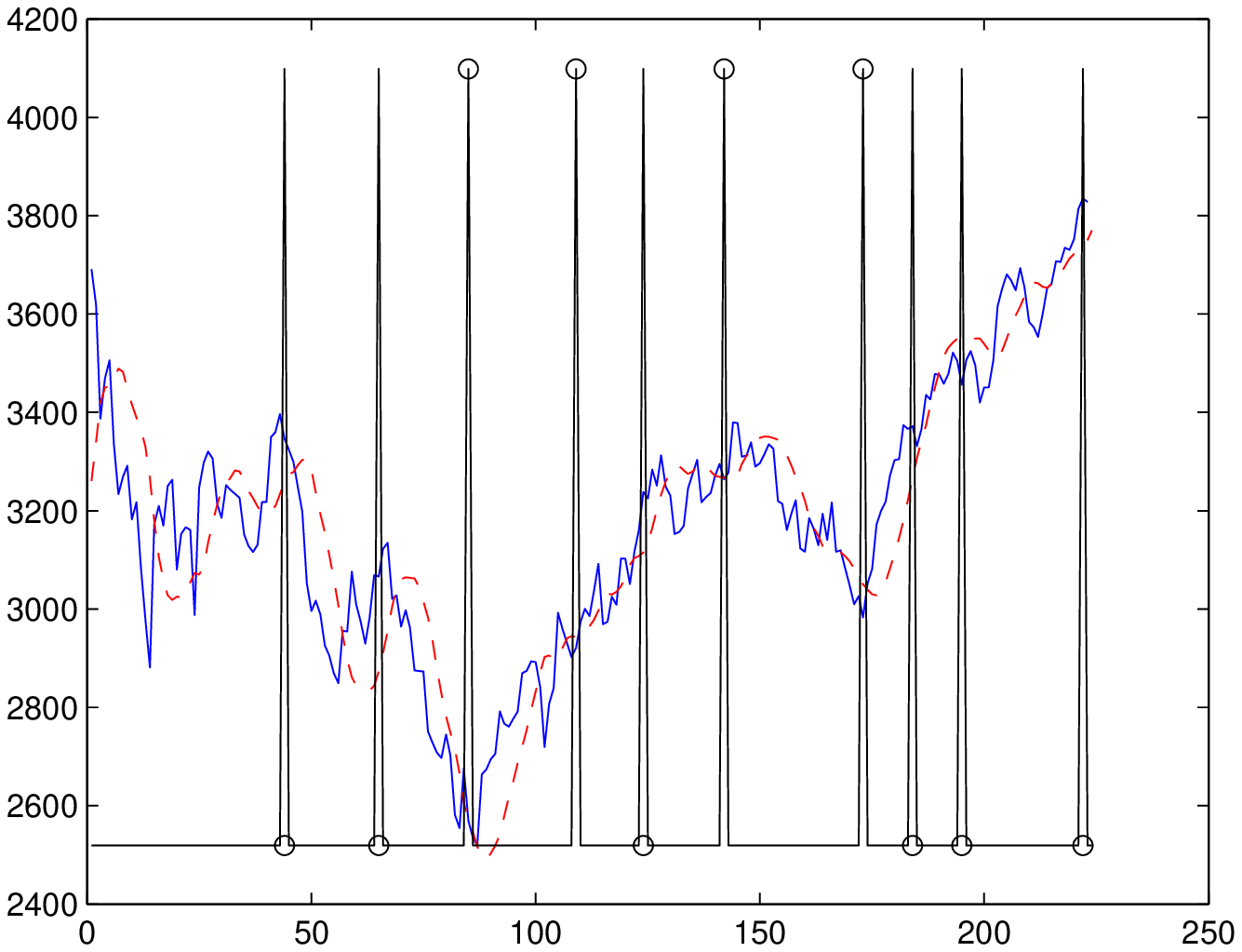}}
\subfigure[\scriptsize Risk-free $\Delta$ tracking (--) and $\Delta$ tracking (- -),
prediction of abrupt change locations (l)]{\includegraphics[scale=.5]{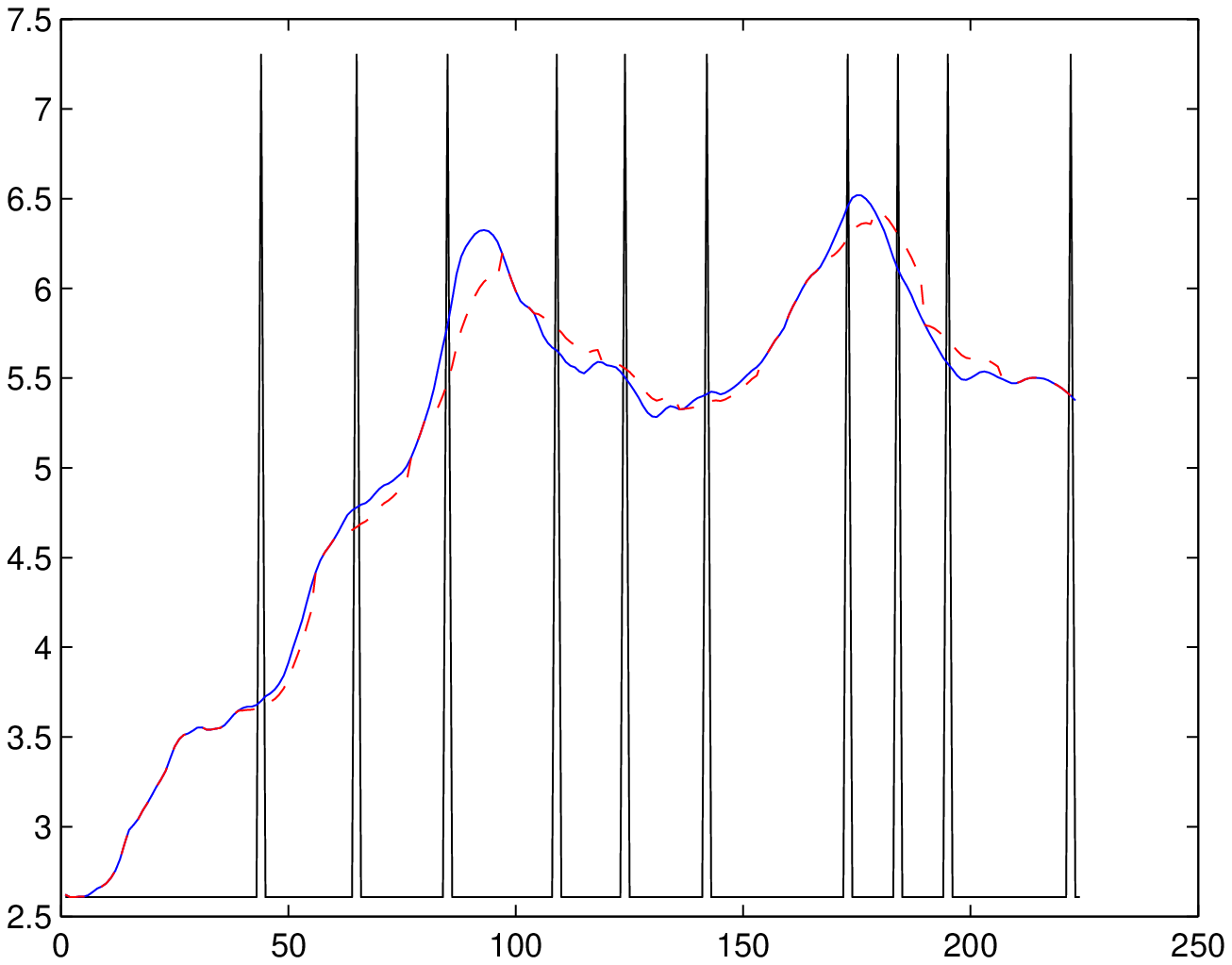}}
\center\subfigure[\scriptsize Zoom on (b)]{\includegraphics[scale=.5]{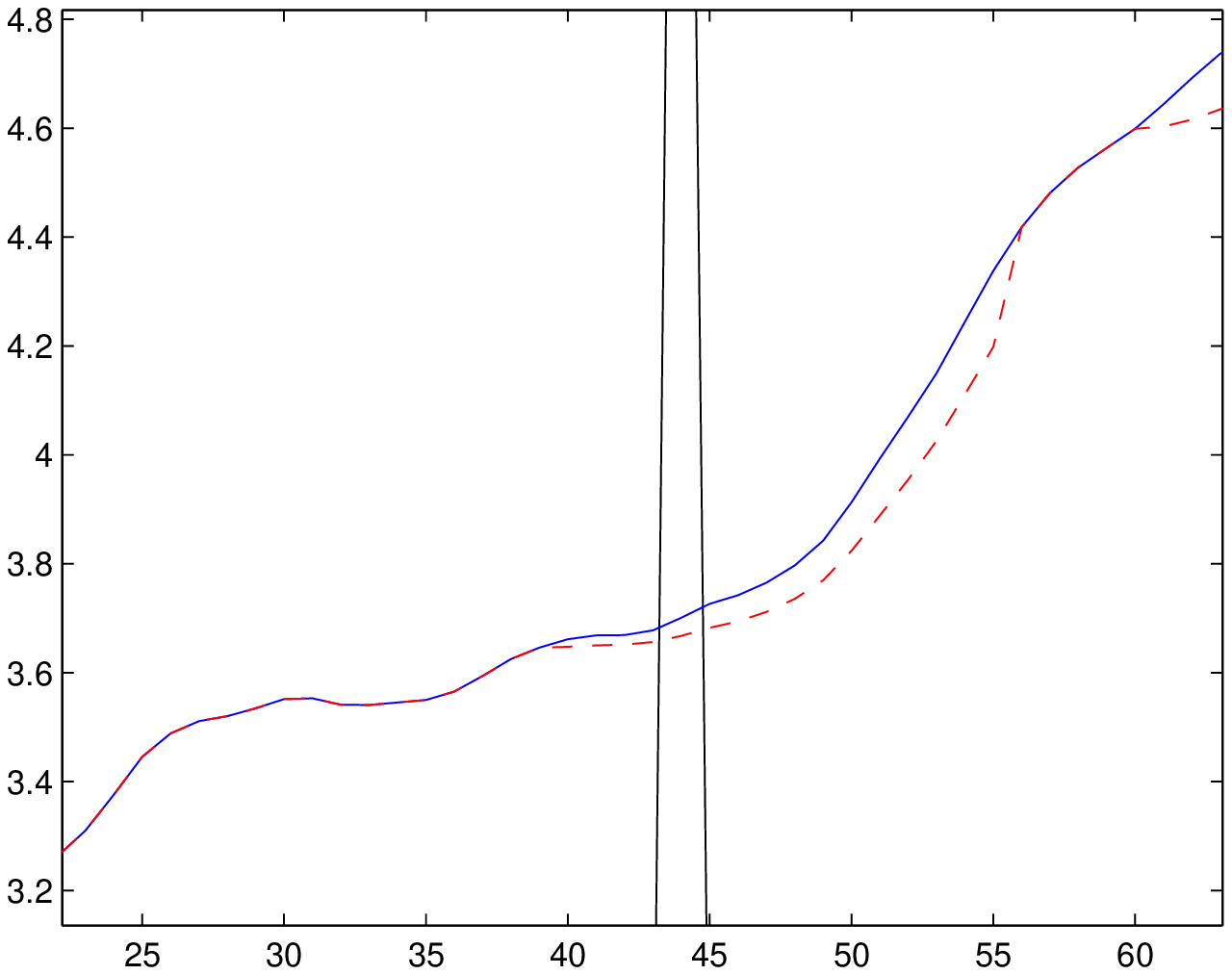}}
\subfigure[\scriptsize Zoom on (b)]{\includegraphics[scale=.5]{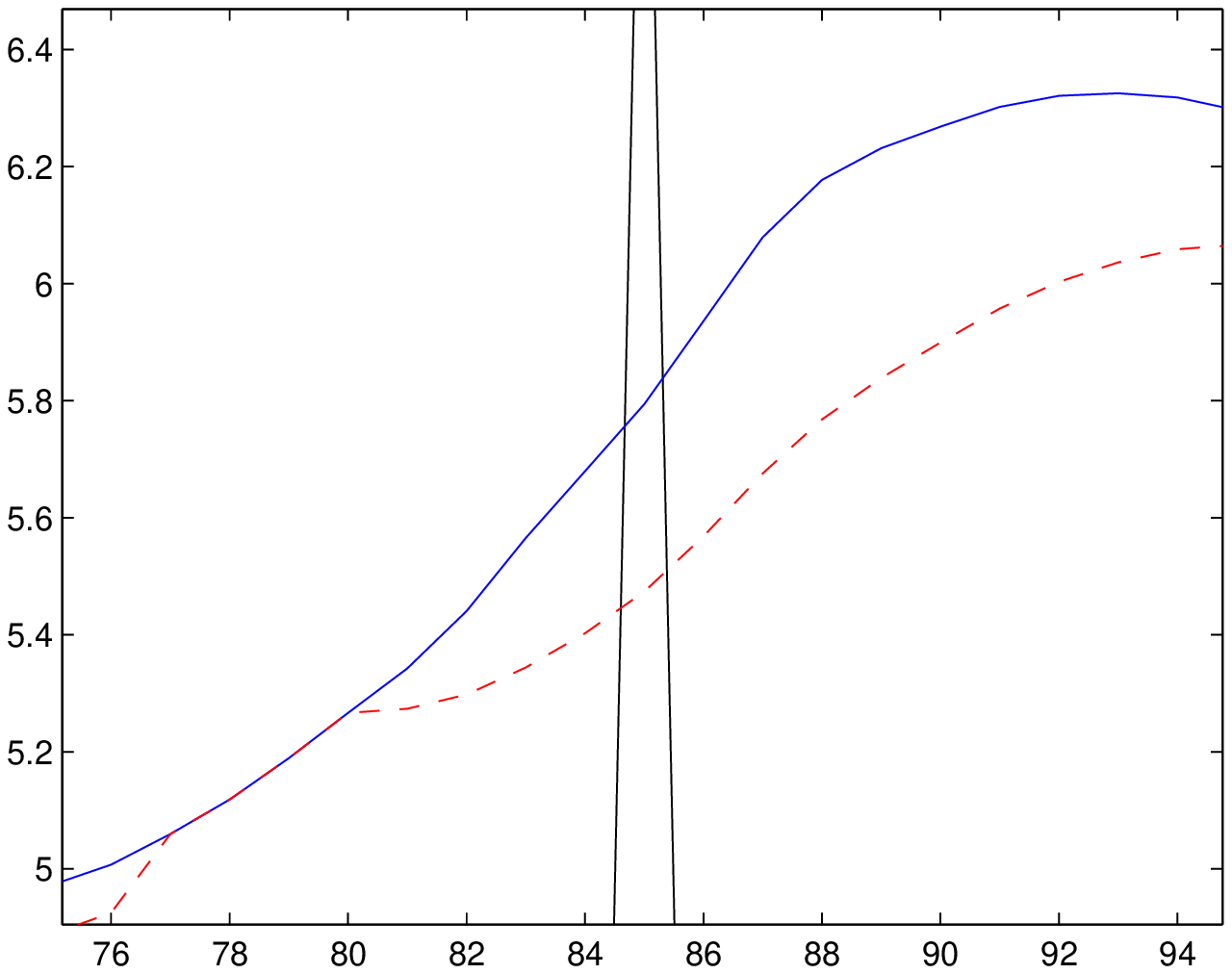}}
\caption{Example 1 (continued): CFU9PY3500 }\label{anti}
\end{figure*}

Taking advantage of the above forecasts allows to avoid the
risk-free tracking strategy \eqref{hedging}, which would imply too
strong variations of $\Delta$ and cause some type of market
illiquidity. The Figures \ref{anti}-(b,c,d) show some preliminary
attempts, where other less ``violent'' open-loop tracking controls
have been selected.
\begin{remark}
Numerous types of dynamic hedging have been suggested in the
literature in the presence of jumps (see, {\it e.g.},
\cite{cont,merton,wilmott} and the references therein). Remember
moreover the well known lack of robustness of the BSM setting with
jumps \cite{el}.
\end{remark}

\section{Conclusion}\label{conclusion}
Lack of space prevented us from
\begin{itemize}
\item examining more involved options, futures, and other
derivatives, than in Section \ref{variant},
\item thorough numerical and experimental comparisons with the BSM delta
hedging.
\end{itemize}
Subsequent works will do that, and also revisit along the same lines
other notions which are related to variances and covariances.

We will
\begin{itemize}
\item not try to replace the Gaussian assumptions by more ``complex''
probabilistic laws,

\item further tackle uncertainty by going deeper into the
Cartier-Perrin theorem \cite{cartier}, {\it i.e.}, via a renewed
approach of time series.
\end{itemize}
This is an extreme departure from most today's criticisms of
mathematical finance.

\vspace{1.9cm} \noindent{\bf Acknowledgement}. The authors would
like to thank Fr\'{e}d\'{e}ric {\sc Hatt} ({\it Mereor Investment Management
and Adivisory, SAS}) for stimulating discussions.


\end{document}